# Calculating the optimum pressure and temperature for vacancy minimization from theory; Niobium is an example

Jozsef Garai

*Department of Mechanical and Materials Engineering, Florida International University, Miami, USA*

Self-resonance in the atomic vibration occurs when the average wavelength of the phonon thermal vibration is equivalent or harmonic of the diameters of the atoms. It is suggested that applying pressure at temperature corresponding to the self-resonance should effectively reduce the number of vacancies. This theoretical prediction is tested on Niobium by measuring the magnetic susceptibility of the untreated and treated samples. The applied pressure-temperature treatment increased the critical temperature of Niobium by about 30 percent which was also accompanied with volume increase.

The reduction of the number of vacancy in a substance improves the physical properties of the materials. Traditionally pressure and temperature treatments are employed to achieve this goal. The optimum treatments are usually determined by try and error method. Following a theoretical approach and using conventional thermodynamic relationships equations allow to calculate the optimum pressure and temperature for vacancy minimization are derived here.

When the wavelength of the average thermal phonon vibration is equal to or harmonic with the atomic diameter then self resonance occurs. It is suggested that the application of pressure at the temperature corresponding to self-resonance should result in the reduction of the number of vacancies. The same procedure can also be used to enhance diffusion. Using fundamental thermodynamic relationships the pressure-temperature curve for self-resonance is derived here.

The average wavelength $[\bar{\lambda}]$ of the phonon frequency at a given temperature can be calculated[1] as:

$$\bar{\lambda} = \frac{h\upsilon_B}{k_B T} \qquad (1)$$

where h is the Planck constant, $k_B$ is the Boltzmann constant, $\upsilon_B$ is the bulk seismic velocity and T is the temperature. The bulk seismic velocity is calculated as:

$$\upsilon_B = \sqrt{\frac{K_T}{\rho}} = \sqrt{\frac{VK_T}{N_A M}}, \qquad (2)$$

where $K_T$ is the isothermal bulk modulus, $\rho$ is the density, V is the volume, $N_A$ is the Avogadro's number and M is the mass of the atom. In Eq. (2) the isothermal bulk modulus is used instead of the adiabatic. The difference between the two bulk modulus is usually under 1% which is ignored in this study.

The effect of temperature on the bulk modulus at 1 bar pressure is calculated[2] as:

$$K_{0T} = K_o e^{-\int_{T=0}^{T} \alpha_T \delta dT} \qquad (3)$$

where $K_o$ is the bulk modulus at zero pressure and temperature, $\alpha_T$ is the volume coefficient of thermal expansion and $\delta$ is the Anderson- Grüneisen parameter, which defined as:

$$\delta \equiv \left(\frac{\partial \ln K_T}{\partial \ln V}\right)_p = -\frac{1}{\alpha_{V_p}}\left(\frac{\partial \ln K_T}{\partial T}\right)_p = -\frac{1}{\alpha_{V_p} K_T}\left(\frac{\partial K_T}{\partial T}\right)_p. \qquad (4)$$

Assuming that the temperature and pressure effect on the volume coefficient of thermal expansion and bulk modulus is linear respectively then Eq. (3) can be written as:

$$K_{p,T} = (K_o + K'_o p) e^{-(\alpha_o + \alpha_1 T)\delta T} \qquad (5)$$

where $K'_o$ is the linear term for the pressure dependence of the bulk modulus, $\alpha_o$ is the projected value of the volume coefficient of thermal expansion at zero pressure and temperature and $\alpha_1$ is the linear term for the volume coefficient of thermal expansion. Eventhough, the temperature dependence of the coefficient below the Debye temperature is not linear[3], the linear approximation and Eq. (5) can be used for ambient conditions and at higher temperatures[4] because the introduced error is minor.

The molar volume of the solid at the temperature of interest is calculated by using the EoS of Garai[4], which is given as:

$$V = nV_o e^{\frac{-P}{K_0 + K_{1P}P + K_{2P}P^2} + (\alpha_o + \alpha_{1P}P + \alpha_{2P}P^2)T + \left(1 + \frac{\alpha_{1P}P + \alpha_{2P}P^2}{\alpha_o}\right)^a \alpha_{1T}T^2} \qquad (6)$$

where, $K_{1P}$ a is a linear, $K_{2P}$ is a quadratic term for the pressure dependence of the bulk modulus, $\alpha_{1P}$ is a linear and $\alpha_{2P}$ is a quadratic term for the pressure dependence of the volume coefficient of thermal expansion, $\alpha_{1T}$ is a linear term for the temperature dependence of the volume coefficient of thermal expansion and a is constant, characteristic of the substance. The theoretical explanations for Eq. (6) and the physics of the parameters are discussed in detail[4].

The atomic diameter [d] corresponding to the size of the vacancy is approximated as:

$$d = \sqrt[3]{\frac{V(T,p)}{N_A}}. \qquad (7)$$

Self resonance can occur when

$$\frac{n}{2}\overline{\lambda} = d(T,p) \qquad \text{where} \qquad n \in \mathbb{N}^*. \tag{8}$$

If the thermodynamic parameters are available then the p-T curve for the atomic self-vibration can be calculated by using Eq. (1)-(8).

The vibrational motion of crystals is very complex and idealized approach is valid only for monoatomic highly symmetrical atomic arrangement. Niobium, which satisfies this criterion, has been selected to test the proposed hypothesis because of its importance in superconductivity.

The thermodynamic parameters required for Eqs. (1)-(8) are determined by unrestricted fitting using the available experimental data[5-7]. The 77 experiments cover the temperature and pressure range 293-2470 K and 0-134 GPa respectively. The determined parameters are given in Table 1. The temperatures for the fundamental and the second harmonics at atmospheric pressure and at 10 GPa are calculated by using Eqs. (1)-(8). Using atomic radius 1.46 Å for Niobium[8] the temperatures for the first harmonic at atmospheric pressure and 10 GPa are 340 K and 360 K respectively while for the second harmonic the temperatures are 640 K and 681 K.

The ¼" x 0.5 mm Nb (99.95%) samples were bought from Smart Elements. The approximately 0.1 GPa pressure was achieved by screws in a pressure vessel. The sample under pressure was heated up from room temperature to the targeted temperature (640 K) in 15 hours. The sample was annealed at 640 K for 8 hours and cool down to room temperature in 15 hours. The sample was taken out from the pressure vessel right before the experiment. The effectiveness of the treatment was tested by measuring the critical temperature of the untreated and the treated samples.

The temperature dependence of the ac susceptibility from the treated and untreated Nb sample, performed by using a mutual inductance technique at an applied field of H = 20 Oe and frequency of f = 1 kHz is shown in Figure 1. The real part, $\chi'$, reveals a large diamagnetic signal below 6.8 K and 8.8 K marking the superconducting transition for the untreated and treated sample respectively. Below 6.0 K and 8.0 K, $\chi'$ is flat, indicating that the superconducting transition is complete for both the untreated and the treated samples[9]. The critical temperature of Nb reported in the literature for untreated sample is lower[10] then 6.8 K measured here. The most likely explanation for this lower critical temperature is the presence of iron in the Nb sample. Comparing the untreated and treated samples it is evident that the treatment increased both the critical temperature and the volume by about 30% (Fig. 1). It is concluded that the presented theoretical approach can successfully applied to calculate the optimum pressure and temperature conditions for vacancy minimization.

**Acknowledgement:** I would like to thank Rongying Jin for measuring the magnetic susceptibility and Andriy Durygin for helping in the design and the manufacturing of the pressure vessel.

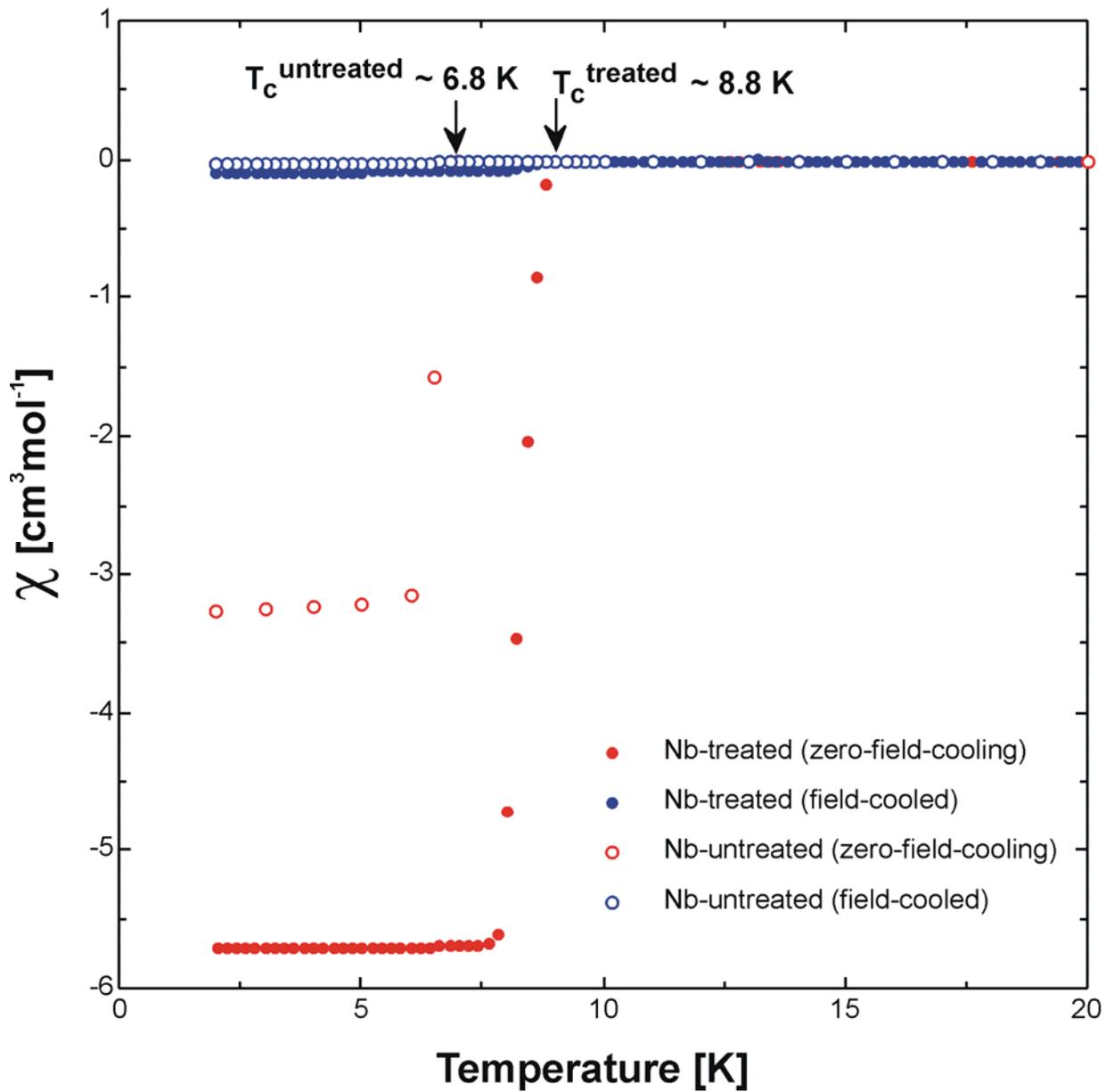

**FIG. 1** Magnetic measurements of the treated and untreated samples[9].

**TABLE 1.** Thermodynamic parameters for the Equation of States.

| 6-306 GPa 293-1380 K [N=334] | $V_o^m$ [cm$^3$] | $K_o$ [GPa] | $K_0'$ | $\alpha_o$ [10$^{-5}$K$^{-1}$] | $\alpha_1$ [10$^{-9}$K$^{-2}$] | $K_{1P}$ | $K_{2P}$ [×10$^{-3}$ GPa$^{-1}$] | $\alpha_{1P}$ [×10$^{-7}$ GPa$^{-1}$K$^{-1}$] | $\alpha_{2P}$ [×10$^{-10}$ GPa$^{-2}$K$^{-1}$] | a | δ | RMS misfit |
|---|---|---|---|---|---|---|---|---|---|---|---|---|
| V(p,T) (G) | 10.736 | 182.23 | | 2.512 | | 1.137 | 0.074 | -18.16 | 0.881 | 12.0 | | 0.0288 cm$^3$ |
| P(V,T) (B-M) | 10.789 | 177.88 | 3.570 | 1.850 | 2.961 | | | | | | 20.03 | 0.96 GPa |

B-M = Universal Birch-Murnaghan EoS
G = EoS of Garai (2007)
6